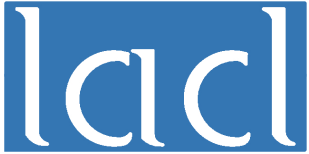
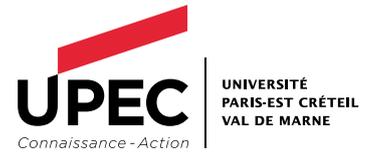

# DBGen User Manual

Emmanuel Polonowski

Laboratory of Algorithmics, Complexity and Logic (LACL)

University Paris-Est Créteil

Technical Report **TR–LACL–2012–4**

Emmanuel Polonowski.

*DBGen User Manual*

# DBGen User Manual

BY EMMANUEL POLONOWSKI

LACL, University Paris-East Créteil

*Email:* emmanuel.polonowski@u-pec.fr

## 1 Introduction

DBGen is a tool for the Coq Proof Assistant. It generates Coq definitions and properties from a term structure with binding handling, providing a framework in the De Bruijn setting.

## 2 Source syntax

The source syntax of DBGen is a valid coq (inductive) term structure annotated with comments for binding information, defined inside a module whose name will be used in the generated content. We assume the knowledge of basic inductive definitions in Coq and of De Bruijn encodings. Annotations are given as Coq comments, *i.e.* between (* and *), placed in strategic locations to indicate the binding structure of the defined syntax.

**Example 1.** Here follows the source syntax for ordinary $\lambda$-calculus:

```
Module LambdaTerms.

  Inductive term : Type :=
  | var ((* index *) x : nat)
  | app (t1 : term) (t2 : term)
  | lam ((* bind term in *) t : term).

End LambdaTerms.
```

### 2.1 `index` annotation

Given a term structure, the (* index *) annotation must be placed before every index arguments in any syntactic category. It will generate an index structure for the category, along with lifting and substitution functions able to deal with it.

**Remark 2.** Indices definition are subject to the following restrictions:

- Constructors having index parameters (as var in Example 1) may not have other arguments.
- Only one index constructor per category is allowed.

### 2.2 `bind` annotation

Binding annotations are placed before the subterm in which the binding occurs. A binding is defined by two informations: the index category that is bound in the subterm and the number of bound variables. Shortcut can be used, as in Example 1, when only a single variable is bound.

**Remark 3.**

- Due to the lack of source code static analysis by DBGen, some errors can occur in the generated code when incorrect binding information is provided. More precisely, there is no verification of the index category name given by the user, and the generated source code will not display informative error message in that case.

- The number of bound variables can be a natural number constant, an identifier (*i.e.* previously defined or declared as a preceding parameter in the same constructor), or an arithmetic expression combining constants and identifiers with the coq standard operations `+`, `-` and `*`. General expression should be allowed as soon as the developer team will embed the coq expressions parser in the tool.

## 2.3 Complete BNF grammar

Here follows the complete source grammar definition.

$$
\begin{array}{rcl}
mod & ::= & \texttt{Module } ModuleName\texttt{.} \\
& & \quad nodes \\
& & \texttt{End } ModuleName\texttt{.} \\[4pt]
node & ::= & \texttt{Inductive } cats\texttt{.} \\[4pt]
cat & ::= & CatName \texttt{ : Type := } constrs \\
& | & CatName \texttt{ : Type := } constrs \texttt{ with } cat \\[4pt]
constr & ::= & \texttt{| } ConstrName\ params \\[4pt]
param & ::= & \texttt{((* index *) } ParamName \texttt{ : nat)} \\
& | & \texttt{((* bind } shifts \texttt{ in *) } ParamName \texttt{ : } CatName\texttt{)} \\
& | & \texttt{(} ParamName \texttt{ : } CatName\texttt{)} \\[4pt]
shifts & ::= & shift \texttt{ , ... , } shift \\[4pt]
shift & ::= & CatName \\
& | & \texttt{[ } exp\ CatName \texttt{ ]} \\[4pt]
exp & ::= & n \\
& | & Id \\
& | & exp \texttt{ + } exp \\
& | & exp \texttt{ - } exp \\
& | & exp \texttt{ * } exp \\
\end{array}
$$

# 3 Generated code

The output of DBGen is a single file defining a module whose name is exactly the module name given in the source file. This allows the user to take advantage of the separate compilation process of Coq.

The generated module is organized as follows:

```
Module ModuleName.

  -- Database and tactics definition.

  -- De Bruijn structure definition.

  -- Lifting and substitution function definitions.

  -- Auxiliary structure and function definitions.

  -- Basic functions properties w.r.t. index cases.
```

```
  -- Index tactic definition.

  -- Advanced functions properties and corresponding tactics.

  -- Main tactic definition.

End ModuleName.
```

The generated Coq code is quite clear and well presented, this allows the user to check and look for definitions and properties he needs in addition to this short (and incomplete) guide.

## 3.1 Name of generated definitions and functions

The initial De Bruijn structure is defined exactly as in the source code: category, constructor and parameter names will be identical. DBGen generate also a named version of the syntax (with strings constants for variable and explicit binding): a '_' is put as a prefix of every names in order to distinguish them from the De Bruijn structure.

For each generated function, a name is build using the names of the involved syntactic categories in order to automatically generate tactics working with those functions. The process of function name generation might help the end user to easily access the function needed in its own Coq development.

Let us consider the following example (formalization of System F):

```
Module SYS_F_terms.

  Inductive type : Type :=
  | tvar ((* index *) i : nat)
  | tconst (n : nat)
  | tarrow (A : type) (B : type)
  | tall ((* bind type in *) A : type).

  Inductive term : Type :=
  | var ((* index *) x : nat)
  | app (t1 : term) (t2 : term)
  | lam (A : type) ((* bind term in *) t : term)
  | tapp (t : term) (A : type)
  | gen ((* bind type in *) t : term).

End SYS_F_terms.
```

Given such a grammar with syntactic categories $\tau_1, ..., \tau_n$, among which $\tau_{i_1}, ..., \tau_{i_k}$ contain an index constructor, DBGen will produce, for every $\tau_{i_m}$ a lifting and a substitution function for every categories $\tau_p$ from which the grammar graph allows to reach $\tau_{i_m}$ (including $\tau_{i_m}$ itself). Hence, for a given pair $(\tau_{i_m}, \tau_p)$, the function name will be build from the names of those categories.

- Lifting function name: $\tau_{i_m}\_\texttt{lift\_in}\_\tau_p$, its type will be `nat -> nat -> `$\tau_p$` -> `$\tau_p$.
  For our example, it gives us three functions: `type_lift_in_type`, `type_lift_in_term` and `term_lift_in_term`.

- Substitution function name: $\tau_{i_m}\_\texttt{subst\_in}\_\tau_p$, its type will be $\tau_{i_m}$` -> nat -> `$\tau_p$` -> `$\tau_p$.
  For our example, it gives us three functions: `type_subst_in_type`, `type_subst_in_term` and `term_subst_in_term`.

The developer can use those function, for instance, to define $\beta$-reduction (as a predicate `reduce`):

```
      forall t u : term, reduce (app (lam t) u) (term_subst_in_term u 0 t)
```

Some functions process the grammar only once, without the need of a specific treatment for every indexed categories, for instance the translation function from the named syntax to the De Bruijn syntax. In such cases, only the $\tau_p$ name is used.

## 3.2  Name of generated infrastructure and tactics

At the beginning of the module, a hint database is declared in order to take advantage of coq automation. Its name is build from the module name as follows:
```
Create HintDb ModuleName_database.
```

A tactic is then immediately defined to use this database and perform trivial simplifications and arithmetic proofs (using the library Omega), named `crush_tac`. This tactic is intended to be quick in its work. Another tactic, named `ecrush_tac`, is similar to `crush_tac` but use `eauto` instead of `auto`.

Several other tactics are the defined, whose goal is to simplify arbitrary terms containing generated function in order to help the end user to perform subsequent proofs. The main tactic, named `dbgen_tac`, which combines the strength of all the generated tactics, is probably powerful enough to deal with any specific case. The other tactics can of course be invoked separately by the user, their definitions and names are to be found in the generated module.

# 4  Usage

The DBGen tool takes as argument the source file name and the output file name. If a file exists with the given output file name then it will be replaced by the generated file.

```
usage: dbgen [ -version ][ -debug ] in-file out-file
```

The `-version` option causes DBGen to print its version number and immediately exit. The `-debug` option displays informations about the internal treatment for debugging purposes.

## 4.1  Examples

Several examples are provided in the `Test` subdirectory.

# Table of contents